\font\big=cmbx10 scaled \magstep2
\font\bic=cmbx10 scaled \magstep1

\documentstyle[epsf]{lamuphys}
\makeatletter
\let\chapter\hid@chapter
\makeatother
\def\eps@scaling{.95}
\def\epsscale#1{\gdef\eps@scaling{#1}}
\def\plotone#1{\centering \leavevmode
\epsfxsize=\eps@scaling\columnwidth \epsfbox{#1}}
\begin{document}

~~~~~~~~~~~~~~~~~~~~~~~~~~~~~~~
\thispagestyle{empty}
\setcounter{page}{0}

\begin{center}

\vspace{1.5cm}

{\big The Epoch of Major Star Formation}

\vspace{0.5cm}

{\big in High-z Quasar Hosts}

\vspace{2.5cm}

{\bic Y. Taniguchi$^{1,2}$, N. Arimoto$^{3,4}$, T. Murayama$^{1}$,}

\vspace{0.5cm}

{\bic A. S. Evans$^{5,6}$, D. B. Sanders$^{5}$, \& K. Kawara$^{7,8}$}

\vspace{1cm}

{$^1$ Astronomical Institute, Tohoku University, Aoba, Sendai 980-77, Japan}

\vspace{0.3cm}

{$^2$ Royal Greenwich Observatory, Madingley Road, Cambridge CB3 0EZ, UK}

\vspace{0.3cm}

{$^3$ Institute of Astronomy, University of Tokyo, Mitaka 181, Japan}

\vspace{0.3cm}

{$^4$ Physics Department, University of Durham, South Road, Durham, DH1 3LE, UK}

\vspace{0.3cm}

{$^5$ Institute for Astronomy, University of Hawaii, Honolulu HI 96822, U. S. A.}

\vspace{0.3cm}

{$^6$ Astronomy Department, Caltech, Pasadena, CA 91125, U. S. A.}

\vspace{0.3cm}

{$^7$ ISAS, Yoshianodai, Sagamihara 227, Japan}

\vspace{0.3cm}

{$^8$ ISO Science Operations Centre,  Villafranca, Spain}

\vspace{2.5cm}

{\it To appear in the Proceedings of}

{\bf Quasar Hosts}

{held at Tenerife, Spain (September 24-27, 1996)}

\end{center}

\newpage

\pagenumbering{arabic}
\title{The Epoch of Major Star Formation in High-z Quasar Hosts}

\author{Yoshiaki\,Taniguchi\inst{1, 2}, Nobuo\, Arimoto\inst{3,4}, 
Takashi\, Murayama\inst{1}, Aaron S.\, Evans\inst{5,6}, David B.\ Sanders\inst{5},
and Kimiaki\,Kawara\inst{7, 8}}

\institute{Astronomical Institute, Tohoku University, Aoba, Sendai 980-77, Japan
\and
Royal Greenwich Observatory, Madingley Road, Cambridge CB3 0EZ, UK
\and
Institute of Astronomy, University of Tokyo, Mitaka 181, Japan
\and
Physics Department, University of Durham, South Road, Durham, DH1 3LE, UK
\and
Institute for Astronomy, University of Hawaii, Honolulu HI 96822, U. S. A.
\and
Astronomy Department, Caltech, Pasadena, CA 91125, U. S. A.
\and
ISAS, Yoshianodai, Sagamihara 227, Japan
\and
ISO Science Operations Centre,  Villafranca, 
28080 Madrid, Spain}

\authorrunning{Taniguchi et al.}
\maketitle

\begin{abstract}
We present the results of our observing program
on NIR spectroscopy of high-redshift ($z$) quasars which have been undertaken both at
Kitt Peak National Observatory and at Mauna Kea Observatory, University of Hawaii.
These data are utilized for studying the epoch of major star formation in
high-$z$ quasar hosts.
\end{abstract}

\section{Introduction}

The major epoch of star formation in galaxies is one of the
most important topics in modern astrophysics,
because it is significantly related to the formation of galaxies
and quasars as well as to cosmology.
Massive stars formed in the first episode of
star formation have a lifetime of
$10^6$ to $10^7$ years
and then  release Type II supernova (SNII) products
(primarily the $\alpha$-elements such as O, Ne, Mg, Si, etc., but
comparatively little iron).
It takes a much longer time for
Type Ia supernovae (SNIa)
to release  iron.
The different nucleosynthesis yields and timescales of SNIa's
and SNII's thus make the abundance ratio [$\alpha$/Fe]
a potentially useful cosmological clock with which one
can identify the epoch of first star formation in galaxies.
It is therefore important to
study chemical properties of high-redshift ($z$) objects.

Since it is considered that the heavy elements in the broad line regions (BLRs)
come from stars in a host galaxy,
systematic study of chemical properties of BLRs
of quasars at high redshift is of particular interest
(\cite{ham93}).
Rest-frame optical emission lines, which are usually used to study chemical
properties of nearby objects,  are redshifted to the near-infrared (NIR)
in these quasars.
Recent NIR spectroscopy of high-$z$ quasars has shown that
the rest-frame optical spectra are dominated by singly ionized iron (FeII) emission
as well as
hydrogen recombination lines (Hill, Thompson, \& Elston 1993;
Elston, Thompson, \& Hill 1994)
suggesting long-lasting star formation
in the nuclear regions of the quasar hosts ($\sim 1$ Gyr).

In order to study the major epoch of star formation in high-$z$ quasar hosts,
we present new results of our NIR spectroscopy of high-$z$ ($z > 3$)
quasars; 1) B1422+231 ($z=3.62$; \cite{pat92}), 
2) PKS 1937$-$101 ($z = 3.79$; Lanzetta et al. 1991), and
3) S4 0636+68 ($z= 3.2$; Stickel \& Kuhr 1994).

\section{Observational Results}

The two quasars, B1422+231 and PKS 1937$-$101, were
observed by using the long-slit
Cryogenic Spectrometer (CRSP)
 with a $256 \times 256$ InSb detector array at the f/15 focus of
the Kitt Peak National Observatory
 (KPNO) 4 meter telescope
while the other quasar, S4 0636+68, was observed 
by using the KSPEC at the Cassegrain focus of the UH 2.2 m telescope.
The details of the observations and the data reduction are given 
elsewhere (Kawara et al. 1996; Murayama et al. 1997).
The spectra of the three quasars are shown in Fig. 1.
We describe their important observational properties below.

\subsection{B1422+231}

The spectrum in Fig. 1 shows 
the emission lines, MgII$\lambda$2798, H$\gamma$, H$\beta$, and
[OIII] $\lambda 5007$ as well as a marginal detection of
CIII] $\lambda 2326$.
Note that this is the first detection of [OIII]$\lambda 5007$ in a  quasar
beyond $z = 3$ (Kawara et al. 1996).
[OIII]$\lambda 5007$ relative to H$\beta$ is smaller
in B1422+231 than the LBQS composite (Francis et al. 1991). 
The broad feature of optical Fe II emission lines is present .
The feature shortward of MgII emission line
is due to UV Fe II emission lines.
 The flux ratio
Fe II(UV + opt)/Mg II of B1422+231 is comparable to that of the LBQS
composite spectrum: $12.2 \pm 3.9$ for B1422+231 and 8.9 for the LBQS
composite. Note that Fe II(UV) and Fe II(opt) denote Fe II emission in
2000 ${\rm -}$ 3000 \AA\ and 3500 ${\rm -}$ 6000 \AA\ in the rest-frame, respectively.
Wills, Netzer, \& Wills (1985) give a mean 
ratio of $7.8 \pm 2.6$ for nine low-$z$ quasars with $z = 0.15 {\rm -} 0.63$.
It is thus suggested that the major iron enrichment has already been done
in this quasar host.

\subsection{PKS 1937$-$101}

The [OIII]/H$\beta$ ratio is similar to that of LBQS composite quasar spectrum.
Since the observed $K$-band spectrum can be fit well solely by
the emission lines of [OIII]$\lambda$4959,5007, H$\beta$, H$\gamma$,
and the linear continuum,
there seems little
optical FeII emission which is ubiquitously observed
in either  high-$z$ quasars (Hill et al. 1993; Elston et al. 1994; Kawara et al. 1996)
or most low-$z$ quasars (Boroson \& Green 1992).
The $J$-band spectrum shows also
little evidence for UV FeII emission feature, either.
We fit the continuum emission with a power law of 
$F_\nu \propto \nu^{-0.50}$,
which is almost consistent with the average continuum spectrum of quasars,
where the power-law index ranges from $-0.3$
(Francis et al. 1991) to $-0.7$ (Sargent et al. 1989).
The UV spectra of most quasars,
regardless of radio loudness (Bergeron \& Kunth 1984),
are dominated  by the FeII features as well as
the power-law continuum emission.
Therefore, both the lower flux and the featureless property of the $J$ band spectrum
are explained by the absence of UV FeII emission features
in PKS 1937$-$101.
In the red edge of the $J$-band spectrum,
a blue part of MgII$\lambda$2798 emission can be seen.

\subsection{S4 0636+68}

The NIR spectrum of this quasar was first reported by Elston et al. (1994)
who showed that its rest-frame optical spectrum is significantly dominated 
by FeII emission lines, suggesting an  iron overabundance 
than in the solar neighbourhood.
Although our new measurement has confirmed the presence of FeII$\lambda5169$ emission,
its intensity relative to that of H$\beta$ emission
is significantly weaker than that of Elston et al. (1994).
The intensity ratio of FeII(opt)/H$\beta$ is estimated
to be 3.5$\pm$1.1. This value is slightly larger than those of low-$z$ quasars;
1.63$\pm$0.88 (Wills et al. 1985). However, we cannot conclude that 
S4 0636+68 belongs to a class of strong iron  quasars
(cf. L\'ipari, Terlevich, \& Macchetto 1993).

\begin{figure}
\epsscale{0.76}
\plotone{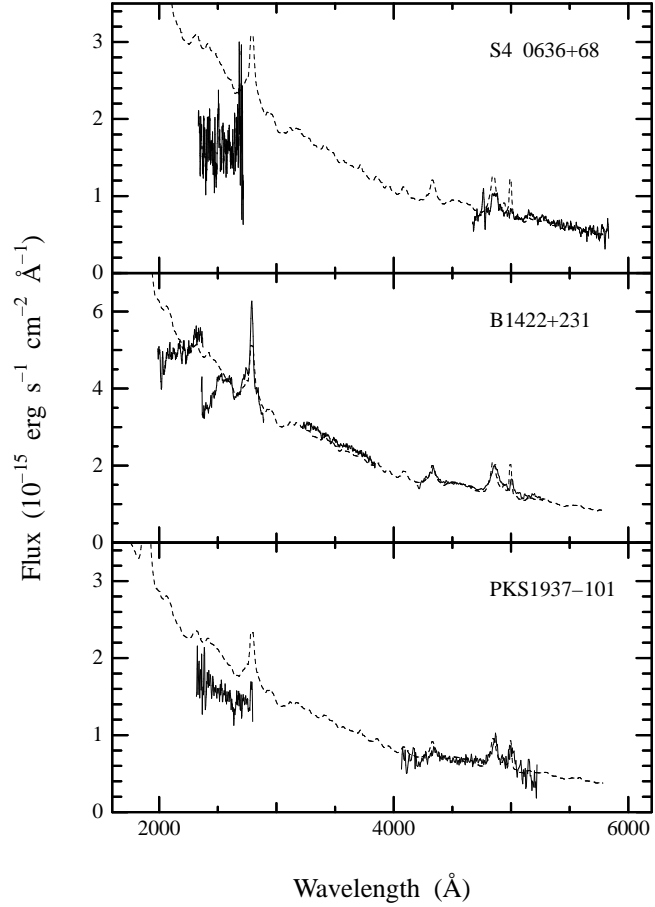}
\caption{The rest-frame optical spectra of the three high-z quasars; S4 0636+68, B1422+231,
and PKS 1937$-$101. The dashed spectrum in each panel is the mean spectrum
of LBQS quasars taken from Francis et al. (1991).}
\end{figure}

\section{Discussion}

We discuss the nature of high-$z$ quasars in viewed from their rest-frame
optical spectra. 
There is a tendency that the quasars with $z < 3.5$ show strong FeII emission
(Hill et al. 1993; Elston et al. 1994)
while those with $z > 3.5$ show strong [OIII] emission.
It should be, however, mentioned that
the strong optical FeII emission of S4 0636+68 reported by Elston et al. (1994)
is not confirmed in this study.
One interesting spectroscopic property known for low-$z$ quasars is
the anticorrelation between
the strength of optical FeII and [OIII] emission lines,
although its physical mechanism is not fully understood (Boroson \& Green 1992).
We examine if the high-$z$ quasars follow the same anticorrelation.
In Fig. 2, we show the relationship of the equivalent width ratios between
([OIII]$\lambda4959+\lambda5007$)/H$\beta$ and FeII$\lambda$4434-4684/H$\beta$.
The low-$z$ quasars studied by Boroson \& Green (1992) show a loose, but
statistically significant  anticorrelation.
It is also known that the radio-loud quasars tend to be located in the lower portion of
this diagram (i.e., weak FeII emitters). PKS 1937$-$101, B1422+231 (Kawara et al. 1996),
and the radio-quiet, high-$z$ quasars studied by Hill et al. (1993)
share the same property
as those of low-$z$ quasars. On the other hand, the radio-loud quasars studied by Elston
et al. (1994) and Hill et al. (1993) do not follow the same as
low-$z$ quasars although our new measurement of 
S4 0636+68 shows that the ratio is consistent with those of low-$z$ quasars. 
If there would be many strong iron radio-loud quasars at high redshifts,
we would have to introduce a new class of quasars which has not yet been
observed at low redshifts.

\begin{figure}
\epsscale{0.68}
\plotone{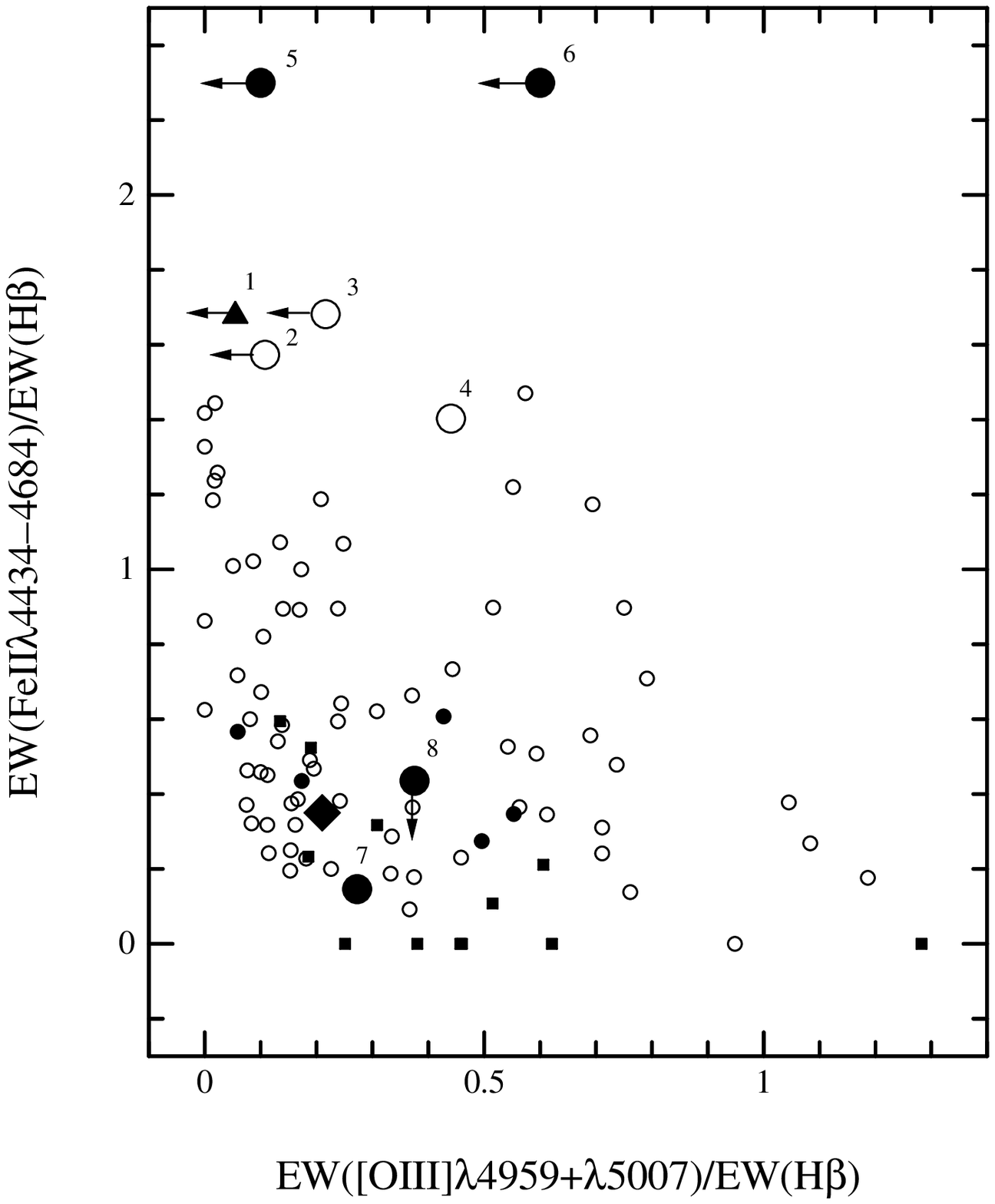}
\caption{
Diagram between ([OIII]$\lambda$5007+$\lambda$4959)/H$\beta$
equivalent width ratio
and FeII$\lambda$4434-4684/H$\beta$ one for low-$z$ (small symbols;
Boroson \& Green 1992)
and high-$z$  ($z >2$) quasars (large symbols; Hill et al. 1993; Elston et al. 1994;
Kawara et al. 1996; this study).
Radio-quiet, radio-loud with flat spectrum, and radio-loud with steep spectrum
are shown by open circles, filled circles, and filled squares, respectively.
B2 1225+317 is shown by the filled triangle because its radio spectrum is unknown.
The numbers given for the high-$z$ quasars correspond to;
1. B2 1225+317, 2. Q1246$-$057, 3. Q0933+733, 4. Q1413+117, 5. S4 0636+68,
6. Q0014+813, 7, B1422+231, and 8. PKS 1937$-$101.
The filled diamond shows our result for S4 0636+68.}
\end{figure}

Finally, we discuss the epoch of major star formation in the host galaxies of
B1422+231 and PKS 1937$-$101.

1) B1422+231:
We show that 
the ratio of  Fe II/Mg II, including UV Fe II emission lines, in the
broad-line gas of some quasars at $z = 3.6$ is almost identical to those at
the low-redshift quasars.  This may imply that the Fe/Mg abundance  at the
center of some quasar host galaxies did not change after
$z = 3.6$.
It is generally considered that 
Mg is preferentially produced in massive star supernovae (SNe II, Ib, and Ic)
on short time scales ($2 {\rm -} 10$ Myr), while Fe is mainly created by
accreting white dwarf supernovae (SNe Ia) in much longer time scales
($1 {\rm -} 2$ Gyr).
The Fe/Mg abundance ratio should be $1/4 {\rm -} 1/2$
of the solar value until SNe Ia start to produce significant amount of Fe.
When SNe Ia dominate the Fe production, the Fe/Mg abundance increases up
to the values in low-redshift quasars and is kept nearly constant since then.
Although it is not straightforward to derive the Fe/Mg abundance from
the present data, the similarity in Fe II/Mg II between B1422+231 and
low-redshift quasars (and the LBQS composite spectrum) suggests that
the host galaxy of B1422+231 had already been in the late evolutionary phase
of the Fe enrichment at $z = 3.6$.
\cite{yos} derived $\sim$\ 1.5 Gyr for the lifetime of SN Ia
progenitors from the analysis of the O/Fe and Fe/H abundances in solar
neighbourhood stars.
If the Fe enrichment started at 1.5 Gyr after
the onset of the first star formation, the host galaxy of B1422+231 would
have formed at $z \geq 15$  for $q_{0}$ = 0.0 and $H_{0}$ = 100 km s$^{-1}$ Mpc$^{-1}$
while at $z \geq 6$ for $q_{0}$ = 0.0 and $H_{0}$ = 50 km s$^{-1}$ Mpc$^{-1}$.

2) PKS 1937$-$101: 
The little evidence for Fe emission lines suggests that the major
epoch of star formation in this quasar host is different from 
that in B1422+231.
The $\alpha$ elements, such as O and Mg, come from SNII's of massive star origin
and thus are quickly expelled into the interstellar space after
the major episode of star formation
(within a few $10^6$ to 10$^7$ years).
It is considered that the N enrichment is delayed ($\sim 10^8$ years)
because it is partly a secondary element formed by CNO burning
in stellar envelope (Hamann \& Ferland 1993).
The rest-frame ultraviolet spectra of PKS 1937$-$101 taken by
Lanzetta et al. (1991) and Fang \& Crotts (1995)
show evidence for NV$\lambda$1240 emission.
Therefore, the nuclear gas has already been polluted with N,
implying that the elapsed time from the major star formation is longer than
$\sim 10^8$ years (Hamann \& Ferland 1993).
However, our observation has shown that the major Fe enrichment  has not yet been
made in PKS 1937$-$101.
The bulk of iron come from SNIa's whose
progenitors' lifetime is very likely to cluster around $\sim 1.5$ Gyr
(Yoshii et al. 1996).
Therefore, the Fe enrichment may start at 1.5 Gyr after
the onset of the first, major  star formation in quasar host galaxies.
These arguments, therefore,  specify the epoch of major star formation in PKS 1937$-$101;
$\sim 10^8$ - 1.5$\times 10^9$ years before redshift 3.787.
Namely, the initial star formation would occur
at $ 3.9 < z  < 6.7$ for
$H_0 = 50$ km s$^{-1}$ Mpc$^{-1}$ and $q_0 = 0$, while at $4.0 < z < 17$ for
$H_0 = 100$ km s$^{-1}$ Mpc$^{-1}$ and $q_0 = 0$.
Recent theoretical prescription on the star formation at high-$z$ universe suggests
that the major epoch of star formation may occur $z < 5$ although subgalactic structures may
exist even at $z > 10$ (Rees 1996).
Provided that the smaller $H_0$ is more preferable,
the present observation is consistent with this prescription.

%
%

\end{document}